\documentclass[twocolumn,showpacs,preprintnumbers,amsmath,amssymb,prb]{revtex4}
\usepackage{graphicx}
\usepackage{dcolumn}
\usepackage{bm}

\begin{document}

\title{The pseudomorphic to bulk fcc phase transition of 
thin Ni films on Pd(100)}

\author{G.A. Rizzi$^{a}$}
\author{A. Cossaro$^{b}$}
\author{M. Petukhov$^{a}$}
\altaffiliation[Also at ]{IGNP, Russian Research Center, Kurchatov Institute, Moscow, Russia.}
\author{F. Sedona$^{a}$}
\author{G. Granozzi$^{a}$}
\author{F. Bruno$^{b}$}
\author{D. Cvetko$^{b}$}
\altaffiliation[Permanent Address: ]{Department of Physics, University of Ljubljana, Ljubljana, Slovenia.}
\author{A. Morgante$^{b}$}
\altaffiliation[Also at ]{Department of Physics, University of Trieste, Trieste, Italy.}
\author{L. Floreano$^{b}$\footnote{ Corresponding Author:
 Fax: +39-040-226767; \\ E-mail: floreano@tasc.infm.it}}
\affiliation{$^{a}$Dipartimento di Scienze Chimiche and 
INFM research unit, University of Padova, Via Loredan 4, I-35131 
Padova, Italy.}
\affiliation{$^{b}$Laboratorio TASC dell'Istituto Nazionale per la Fisica della
Materia, Basovizza SS-14, Km 163.5, I-34012 Trieste, Italy.}

\date{\today}

\begin{abstract}

We have measured the transformation of pseudomorphic Ni films on 
Pd(100) into their bulk fcc phase as a function of the film thickness.
We made use of x-ray diffraction and x-ray induced photoemission to 
study the evolution of the Ni film and its interface with the substrate.
The growth of a pseudomorphic film with tetragonally strained face 
centered symmetry (fct) has been observed by out-of-plane 
x-ray diffraction up to a maximum thickness of 10 Ni layers (two of them 
intermixed with the substrate), where a new fcc bulk-like phase is 
formed. After the formation of the bulk-like Ni domains,  
we observed the pseudomorphic fct domains to disappear preserving the 
number of layers  and their spacing. The phase transition thus 
proceeds via lateral growth of the bulk-like phase within the pseudomorphic 
one, i.e. the bulk-like fcc domains penetrate down to the substrate when 
formed. This large depth of the walls separating the domains of different 
phases is also indicated by the strong increase of the intermixing at the 
substrate-film interface, which starts at the onset of the transition 
and continues at even larger thickness. The bulk-like fcc phase 
is also slightly strained; its relaxation towards the orthomorphic 
lattice structure proceeds slowly with the film thickness, 
being not yet completed at the maximum thickness  presently studied 
of 30~\AA~ ($\sim$~17 layers). 

\end{abstract}

\maketitle

\section{Introduction}

Heterogeneous epitaxy is a widely exploited technique to fabricate 
artificial materials, since it allows to introduce a controlled 
degree of distortion of the interatomic bond length and orientation, 
which finally allows one to tune the electronic properties of 
nanostructured devices.
In the case of semiconductors, the strain of a film growing 
on a heterogeneous substrate can be retained up to a thickness 
of several hundreds of nanometers (coherent growth). 
Beyond this critical thickness, 
the strain is released through the formation of a pattern of 
misfit dislocations, which propagate from the interface to the 
surface and drive the gradual relaxation (decoherence) of the 
growing film to its bulk structure. 
Thanks to the high degree of long range order, the 
mechanism for semiconductor decoherence can be followed and described 
in much detail.\cite{tsao} 
While a reduced misfit is required for coherent semiconductor growth,
thin artificial phases with a lattice 
structure much different from the bulk one can be stabilized 
for metal film growth.
In the case of oxide substrates, 
the decoherence of metal films through and ordered network of misfit 
dislocations is usually observed when the misfit does not exceed 
$\sim$~10~\%.\cite{renaud} On the other hand, the metal bonding allows 
the strain to be released on a much smaller thickness scale, thus 
reducing the average domain size of the growing film, i.e. the 
probability of detecting any long range order behaviour. 
Things are getting more complicated when metal on 
metal heteroepitaxy is considered, since intermixing phenomena have 
also to be taken into account, which can either favour the 
stabilization of pseudomorphic films via  surfactant effect 
(like for the Fe/Au(100) system\cite{opitz,feau}) or inhibit the coherent 
growth via substrate roughening and alloying (like for the 
Co/Cu(111) system\cite{demiguel}).

In the past ten years, metal heteroepitaxy has been widely 
applied to study magnetic systems, since
the magnetic behavior of metals (both spin orientation 
and magnetic moment)\cite{moroni} and their chemical reactivity can 
be modified by appropriate distortion of their lattice structure. 
These  metastable artificial phases are  
decomposed into their ground structural phase when the films exceed 
a critical thickness of a few layers. 
These transformations are usually observed to be rather sharp and 
occur through a rather abrupt non-diffusive distortion mechanism, which is 
accompanied by a strong morphological reorganization,
like for the much studied Fe/Cu(100) 
system,\cite{kalki,fecu} thus smearing the mechanisms of domain growth.

In fact, most of the studies have been performed by means of LEED, 
ion scattering and STM, i.e. surface sensitive techniques. 
The behavior of the interface and of the layers beneath the surface 
is poorly known and is usually obtained by photoelectron 
diffraction (XPD) experiments at a Synchrotron facility, 
where the photoelectron kinetic energy can be effectively 
tuned to change the penetration depth. On the other hand, 
XPD data require a rather complex analysis, whose reliability 
is hampered when the number of scatterers (penetration depth) 
is increased.\cite{fadley} Grazing incidence x-ray diffraction (GIXRD) is
certainly the best suited technique to study the structure of 
a buried interface and the layered structure of thin 
films.\cite{renaud} 
Recently, a combination of XPD and GIXRD techniques has been 
applied to study the structural dependence on thickness
of Fe films on Cu$_{3}$Au(100) 
from a pseudomorphic phase to the bcc one.\cite{bruno}
The structural transformation has been shown to be 
characterized by phase coexistence over a thickness range 
of a few monolayers.

Hereafter, we have applied GIXRD to study the structure of thin 
Ni films grown on Pd(100). The epitaxial growth of ultrathin Ni films 
on the Pd(100) surface has been recently studied by means of LEED, 
XPS and XPD with a laboratory x-ray source.\cite{rizzi} 
This study has shown that in the early stages of the deposition 
there is the formation of a tetragonally strained face centered (fct) Ni phase with 
the same lateral lattice spacing of the substrate (i.e. 3.89~\AA) and a vertical 
compression of 0.24~\AA. When the film thickness exceeds a critical 
value of approximately 9-12 layers, the film transforms into its
 bulk-like Ni fcc structure (lattice parameter of 3.52~\AA). 
The measured strain of the fct phase is found to be in full 
agreement with the elasticity theory, but some open questions still remain. 
In particular, although capable to measure very nicely the tetragonal 
strain adopted by the Ni film, the XPD data did not yield any clear 
information on how the phase transition develops as the critical 
thickness is approached nor it was straightforward to extract 
information on the film/substrate interface. For all these reasons we decided to 
perform a more detailed study on this system taking advantage of 
the unique possibilities to acquire both XPS, XPD and GIXRD data 
offered by the ALOISA beamline at ELETTRA Synchrotron (Trieste Italy). 

An effective critical thickness for the pseudomorphic growth has been 
found. The transformation into the Ni bulk-like structure is shown to 
involve the interface as well, where the intermixing  
increases. In addition, the number of layers in the 
residual pseudomorphic fct phase does not change during the 
transformation, but simply its domains shrink. The transformation thus 
proceeds via lateral growth of the bulk-like Ni phase.  The 
increase of the intermixing beyond the critical thickness is 
attributed to the lateral propagation of the defects, which drive the  
transformation of the pseudomorphic phase.

\section{Experimental}

Both electron spectroscopy and
X-Ray surface diffraction measurements have been performed in situ at 
the ALOISA beamline, where a wide photon energy range (130-8000~eV), 
coupled to a multiple detection system, is available.\cite{aloisa} 
The UHV experimental chamber
(base pressure in the 10$^{-11}$~mBar range) hosts  hemispherical
electron analyzers and X-Ray detectors. The emission direction from the
sample surface can be freely selected for any orientation of the surface.
The sample is mounted on a
6--degrees--of--freedom manipulator, specially designed to select with great
accuracy (0.01$^{\circ }$) the orientation of the surface with respect 
to the incoming photon beam. The
temperature of the sample, measured by thermocouples, can be varied by
resistive heating and liquid nitrogen cooling. 

\begin{figure}[tbp]
\includegraphics[width=.5\textwidth]{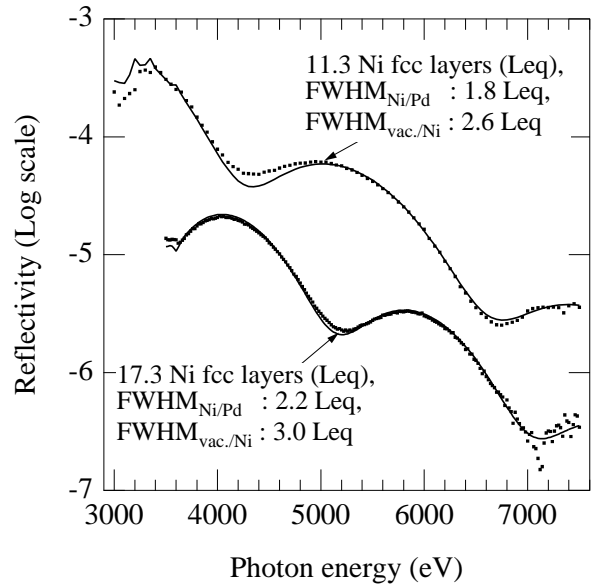}
\caption{\label{fig1}X-ray reflectivity energy scan (dots) taken at a fixed 
scattering angle of 7$^{\circ}$ and 6$^{\circ}$ for two films of 
$\sim$~22 and $\sim$~30~\AA, respectively. The film thickness and 
interface widths are expressed as 
Ni fcc layer-equivalent (Leq) after fitting to the experimental data 
(full lines)
with a simple Ni/Pd bulk model. The Ni/Pd and vacuum/Ni 
interfaces are assumed to have a gaussian depth profile. }
\end{figure}

The Pd(100) substrate was prepared by Ar$^{+}$ sputtering at 1~keV and 
annealing to 970~K. The substrate order was checked by RHEED, while 
XPS surveys at grazing incidence (of the order of the critical
angle) were used to check for residual contamination. 
Nickel was evaporated from a carefully outgassed electron bombardment cell
(Omicron) provided with a water cooled shield. 
A quartz microbalance allowed us to tune the deposition flux
at a constant rate of $\sim$~0.6~\AA/min~ (10~\% accuracy) before deposition.
The absolute calibration of the growth rate was determined a posteriori by measuring 
the x-ray reflectivity (XRR) at fixed scattering angle while scanning 
the photon energy between 3 and 8~keV. The interference between the 
x-ray scattering from the substrate-film and film-vacuum interfaces 
gives rise to maxima and minima as a function of the perpendicular 
momentum transfer. Fitting to the XRR curves with a simple model of 
regularly bulk spaced layers, thus yields the film effective
thickness (from the position of maxima and minima) and the width of 
the interfaces (from the amplitude of the oscillation), as shown in 
Fig.~\ref{fig1}. 

For each Ni film, we measured both in-plane XRD and XPS 
from the Pd 3d core level and the valence band (VB), while XPD polar scans 
from the Ni 2p$_{3/2}$~ were surveyed for a 
better comparison with previous experiments. 
The in-plane XRD measurements consist of radial scans across
the ($\overline{2}$~0~0) peak in the Pd(100) reciprocal lattice.
These measurements were taken scanning the photon energy in a broad range under
a suitable $\theta $-$2\theta $ scattering geometry.  
The observation of diffraction peaks in radial
scans allows us to determine the lateral lattice spacing $d$ through the Bragg
condition $2d\sin \theta =hc/E$.
Out-of-plane 
XRD (rod scans) has been taken for a few selected 
films in the critical thickness range, to determine the 
perpendicular distribution of the Ni layer spacings in the 
pseudomorphic phase. 
The  ($\overline{2}$~0~L) rods of the Pd(100) 
substrate were taken at a photon energy of 7000~eV with a sampling of
$\Delta$L~=0.03, up to L~=2.2. Rod scan simulations were performed by 
the Vlieg's program ROD.\cite{vlieg}

Both Pd 3d and valence band photoemission spectra were taken at a 
photon energy of 500 eV with a photon energy resolution of $\sim 
125$~meV. The surface was kept at a grazing angle of 4$^{\circ}$, 
in transverse magnetic (i.e. $p$) polarization and the electron spectrometer 
was placed along the surface normal with a kinetic energy resolution 
of 170~meV.
XPD polar scans of the Ni 2p$_{3/2}$~ peak (at h$\nu$~=~1270~eV) were
measured in the same scattering conditions by rotating the electron 
analyzer 
in the scattering plane. We considered emission along the two main symmetry
direction $\langle 100\rangle $ and $\langle 110\rangle $ 
of the substrate unit cell. We followed the procedure of 
Ref.~\onlinecite{bruno2} to extract the anisotropy component from the XPD polar 
scans by subtracting an isotropic component, which accounts
for both geometrical (field of view, illuminated area) and physical 
(photoemission matrix symmetry, escape depth, surface roughness)
factors.

\section{Results}

\subsection{Photoemission}

\begin{figure}[tbp]
\includegraphics[width=.5\textwidth]{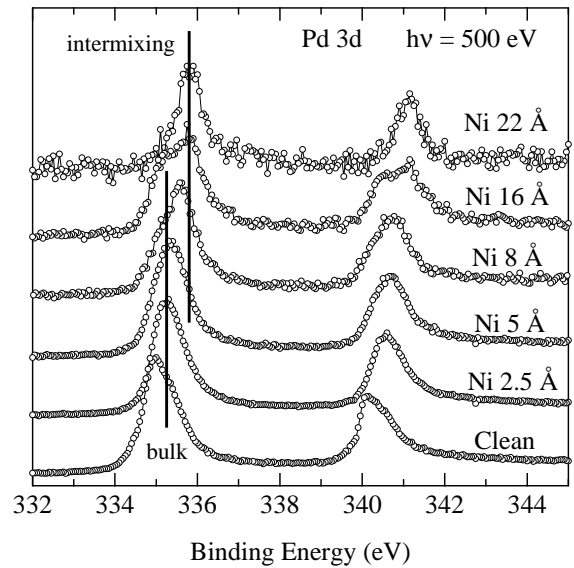}
\caption{\label{fig2} Pd 3d photoemission spectra taken at a photon 
energy of 500~eV with an overall energy resolution of $\sim$~210~meV. 
The bulk component is the shoulder at 335.25~eV binding energy of the Pd 
3d spectrum taken on the clean Pd substrate (bottom curve). The surface 
component yields a core level shift of -0.4~eV to lower binding energy.
As Ni is 
deposited, a new component, due to Pd-Ni alloying, appears 
at a binding energy higher than the 
bulk one. The Pd surface component fully disappears after the 
deposition of 5~\AA~ of Ni. The 
interface component is the main component at 16~\AA~ and beyond.
The vertical bars mark the binding energies of the bulk and interface 
components.
The thickness calibration as been obtained following the procedure 
described in the text and illustrated in Fig.~\ref{fig3}.}
\end{figure}

The Pd 3d XPS data  (with an overall energy resolution of about 
200~meV)
are reported in Fig.~\ref{fig2}. 
The Pd 3d$_{5/2}$ peak of the clean substrate can be fitted with 
two components, one corresponding to the bulk peak (335.25~eV) 
and the other to the surface (334.85~eV) , 
in agreement with previously reported data.\cite{todorova}  
Upon Ni deposition the surface component is replaced by an interface 
one (from 5~\AA~ and beyond), which is shifted
by 1.0 eV (335.8~eV) from the bulk component, towards higher binding energies,  
as can be clearly seen in Fig.~\ref{fig2}, where the bulk 
and interface components are well resolved in the 16~\AA~
thickness film. The bulk component 
disappears after the deposition of 22~\AA~ of Ni. 

\begin{figure}[tbp]
\includegraphics[width=.5\textwidth]{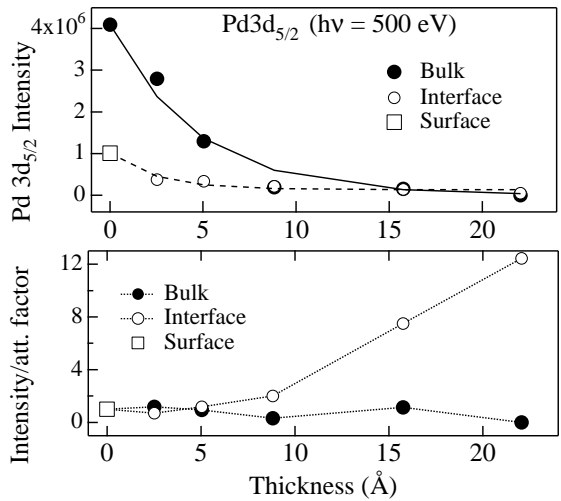}
\caption{\label{fig3} Upper panel: intensity of the Pd 3d$_{5/2}$ 
photoemission peak and its single components (markers) as a function of the 
Ni deposition time. The full line is a 
fit to the bulk component (filled circles) with an exponential decay. 
An additional 
parameter, accounting for intermixing, is needed to fit the interface 
component (dashed line and open circles); in the latter case we used the surface 
component (open square) as the first point of the fitting data.
Lower panel: intensity of the bulk and interface Pd 3d $_{5/2}$ components 
after normalization to the attenuation factor given by the 
exponential decay of the Pd bulk component. The thickness of the Ni film on the 
abscissa axis has been calibrated according to the fitting of the 
Pd bulk attenuation law.}
\end{figure}

The intensity variation of the Pd 3d$_{5/2}$ peak with the film 
thickness is reported in Fig.~\ref{fig3} (upper panel). 
The bulk and interface 
components do not follow the same trend. 
While the bulk component can be nicely fitted with a simple 
exponential decay, where the only fitting parameter is the Ni film 
thickness, i.e. the growth 
rate, the interface component requires a more complex function with an 
extra parameter $\gamma$, which represents the degree of intermixing 
between the film and the substrate. 
It is important to note that the growth rate, 
as determined from the fitting procedure of the bulk component intensity 
(0.63$\pm$0.07 \AA/min), is in full agreement with the one obtained 
from the reflectivity curves. The increasing weight of the interface 
component is well represented in 
Fig.~\ref{fig3}(lower panel), where the ratio between the experimental values 
and the attenuation law, obtained from the bulk component,
is reported. According to the figure, the 
interface component is the dominating one at 16~\AA~ and beyond.

\begin{figure}[tbp]
\includegraphics[width=.5\textwidth]{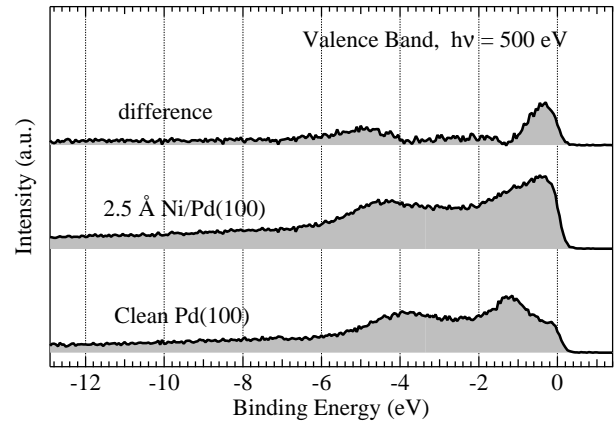}
\caption{\label{fig4}Valence band of the Pd(100) sample before and 
after deposition of 2.5~\AA~ of Ni. The top curve represents the 
difference spectrum of the former ones and puts in evidence the 
satellite peak at -5~eV, which is due to Pd-Ni alloying.}
\end{figure}

The formation of an intermixed phase already at the beginning of the 
deposition is also witnessed by the VB photoemission data of 
Fig.~\ref{fig4}, 
taken just after the deposition of 2.5~\AA~ of Ni. The upper part of 
the figure shows the difference spectra obtained by subtracting the 
clean Pd VB spectrum, multiplied by the proper attenuation factor, 
from the 2.5~\AA~ Ni/Pd(100) spectrum. The result represents the film 
VB, where the peak at a binding energy of 5~eV cannot be found in the 
VB of neither pure Ni(100) nor Pd(100). It is worth to remind that 
the so-called ``6 eV satellite'' of the Ni 2p spectrum is found 
shifted to lower binding energies in the 2.5~\AA~ film with respect to 
the same satellite for a 22~ML thick film.\cite{rizzi} Both the VB 
behaviour and the shift to higher binding energy 
of the Pd 3d$_{5/2}$ core level are the fingerprint of the intermixing 
process at the Ni/Pd interface, i.e. to the formation of a NiPd 
alloy in agreement with literature data.\cite{hillebrecht} 

\begin{figure}[tbp]
\includegraphics[width=.5\textwidth]{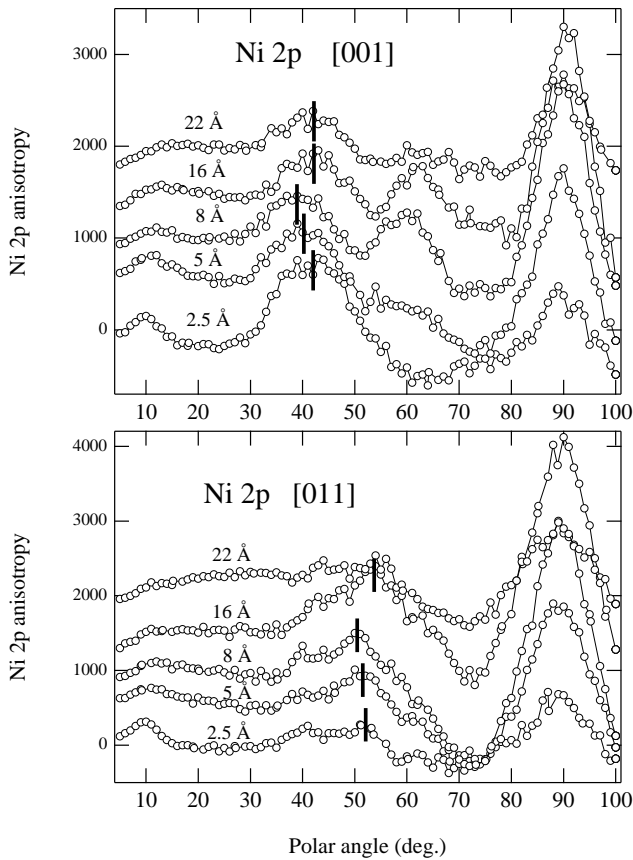}
\caption{\label{fig5} Anisotropy of the polar scans taken for 
the Ni 2p$_{3/2}$~ 
photoelectron peak (kinetic energy of $\sim$~415~eV) 
along the $\left\langle100\right\rangle$ and
 $\left\langle011\right\rangle$ substrate symmetry 
directions, upper and lower panel, respectively. The vertical bars 
indicate the angular position of the forward scattering focussing 
peaks which are characteristic of an fcc symmetry. Deviations from the 
nominal values, i.e. 45$^{\circ}$ 
and 54.7$^{\circ}$ from the surface, are mainly due to distortions of the 
lattice cell in the topmost layers (which changes with the film 
thickness).}
\end{figure}

From the structural point of view, the intermixing at the interface 
determines an enhanced pseudomorphism at the early stages of Ni 
deposition. This can be observed from the thickness dependence of the 
forward scattering peaks originated by focussing from close-compact 
atom row directions in Ni 2p$_{3/2}$~ XPD polar scans, as shown in 
Fig.~\ref{fig5}. At 2.5~\AA, the fcc 
characteristic peaks are 
closer  to the nominal fcc position (at 45$^{\circ}$ and 54.7$^{\circ}$ from 
the surface normal)  than at higher thickness. By 
increasing the film thickness, the forward scattering peaks shift 
away from the surface normal, indicating an increase of the 
vertical compression due to the 
elastic strain. The strain is released beyond 16~\AA~ by the 
transformation into the bulk-like Ni structure, in full agreement with 
previous XPD studies.\cite{rizzi}

\subsection{X-ray diffraction}

\begin{figure}[tbp]
\includegraphics[width=.5\textwidth]{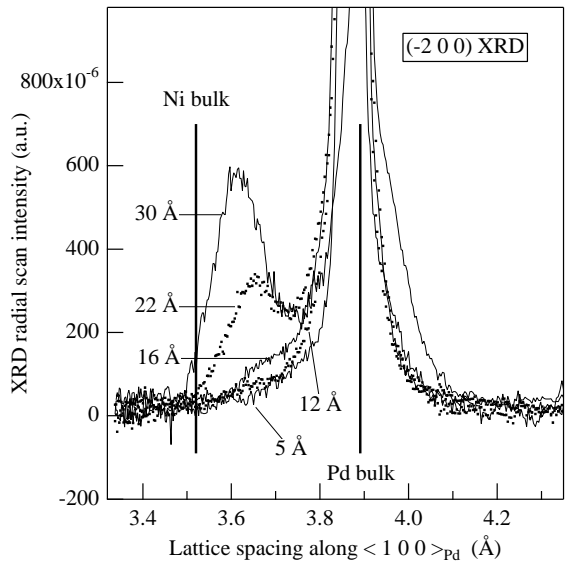}
\caption{\label{fig6} Radial scans of the ($\overline{2}$~0~0) 
XRD peak for a few sample Ni thickness (alternative full and dotted 
lines). The XRD measurements have been taken by scanning the 
photon energy between 5400 and 7000~eV 
at fixed scattering geometry. The photon beam
impinges the surface at grazing incidence, forming an angle 
$\theta \sim 35^{\circ }$ with respect to the fcc(100) planes of 
the direct lattice. The substrate XRD peak is always out of scale, and 
the corresponding lattice spacing is indicated by the vertical thick 
line, as well as the lattice spacing of bulk fcc Ni.}
\end{figure}

The Ni film transformation from the pseudomorphic fct phase to the bulk-like one 
does not occur through a continuous relaxation of the strained lattice 
cell, rather domains of bulk-like symmetry are formed beyond 16~\AA, which 
grow in size as the thickness is increased. This transformation is 
clearly seen in Fig.~\ref{fig6}, where in-plane radial scans across the substrate 
($\overline{2}$~0~0) XRD peak are shown for a few film thicknesses.
Besides the substrate peak, a new feature appears at a thickness of 
16~\AA, corresponding to a lattice spacing of $\sim 3.65$~\AA. 
This feature evolves 
shortly into a well defined peak (see scan at 22~\AA) which gradually shifts 
towards the  3.52~\AA~ lattice parameter of fcc Ni.
The appearance of the Ni bulk-like peak at $\sim 3.65$~\AA~ is the 
fingerprint of a  structural phase transition.

\begin{figure}[tbp]
\includegraphics[width=.4\textwidth]{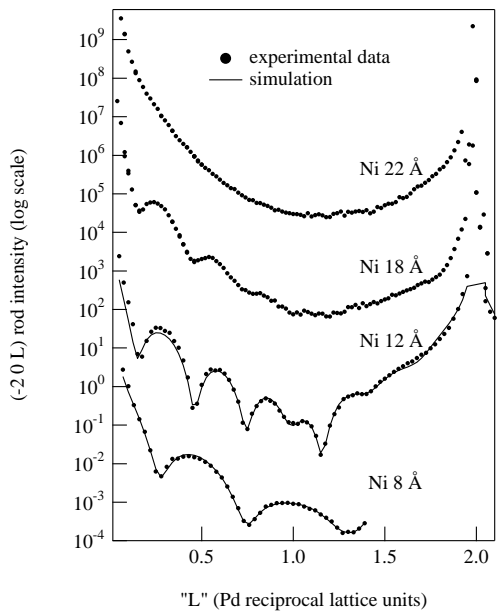}
\caption{\label{fig7} XRD scans of the ($\overline{2}$~0~L) rods of Pd 
for a few Ni films of different thickness. 
Modulations in the low thickness films arise from Bragg interference 
among the Ni pseudomorphic layers, which are confined between 
sharp interfaces at both top and bottom of the film. 
Fitting to the data (full lines) yields a structural model as 
reported in Table~\ref{table1}. }
\end{figure}

For a better understanding of the structural transformation, we have 
studied the layer spacing distribution in the pseudomorphic fct Ni film. 
The XRD ($\overline{2}$~0~L) rod scans are shown in Fig.~\ref{fig7} for 
a few films in the range of the critical thickness of 12-16~\AA. The XRD scan 
along the substrate peak rod puts in evidence the x-ray scattering 
interference between the Ni pseudomorphic layers which are confined 
between the film-substrate and film-vacuum interfaces.
If the deposition is homogeneous on the substrate, well defined 
interference oscillations can be seen along the rod scans, as in the 
case of the 8 and 12~\AA~ film thickness. 
The number of maxima and minima and their position is related to the 
number of Ni layers in the pseudomorphic structure, while the 
amplitude of the oscillations to the sharpness of the Ni film 
interfaces (the larger is the amplitude the sharper is the interface).  
The increased number of maxima and minima from the 8~\AA~ film to the 
12~\AA~ one is thus reflecting the increased number of Ni layers in 
the pseudomorphic structure. The rod scan taken at 
18~\AA, where the Ni film structural transition has already started, is 
still showing a few faint modulations indicating that residual 
domains of the pseudomorphic phase are still co-existing with the fcc 
Ni phase. Most strikingly, the maxima and minima of the 18~\AA~ phase 
still occur at the same perpendicular momentum transfer of the 
12~\AA~ film, i.e. the residual domains of the pseudomorphic phase 
contain the same number of Ni layers. The strong damping of the 
maxima and minima amplitude indicates both an increase of the 
pseudomorphic fct phase 
roughness and the deprival of the pseudomorphic Ni layers 
due to the shrinking of the corresponding domains.
At the thickness of 22~\AA, the whole film has been transformed into 
the Ni bulk-like phase and the rod scan displays a structureless smooth 
behavior, which is characteristic  of a bulk-truncated crystal surface.

The layer spacings and fillings for the 8 and 12~\AA~ films have been 
obtained by fitting the rod scans with the structural model 
reported in Table~\ref{table1}. Due to the large number of fitting 
parameters, we have fixed the inner layer spacings to the value 
determined by PED analysis after Ref.~\onlinecite{rizzi}, 
and we focussed our attention to the substrate-film and film-vacuum 
interfaces, where XRD yields the best sensitivity.
With these constraints, we have found that the topmost layer is also 
compressed, in excellent agreement with previous XPD 
studies,\cite{rizzi} while the layers close to the film-substrate interface are 
expanded, confirming the vertical pseudomorphism indicated by 
the present XPD scan on very thin Ni films (Fig.~\ref{fig5}).
Concerning the buried interface, we admitted Ni/Pd intermixing for two 
layers. While this model yields an excellent fitting for the thinnest film, a 
larger number of layers might be affected by the intermixing in the 
12~\AA~ film, as indicated by the incomplete layer filling of a few 
buried Ni layers (3rd to 5th) with respect to the intermediate ones.
Alternatively, this finding could be the fingerprint of the formation 
of extended morphological defects at the Ni-Pd interface, when the critical 
thickness is achieved.

\begin{table}
\caption{\label{table1} Structural parameters used to fit the 8 and 
12~\AA~ rod scans. The indetermination on the layer spacing and 
fillings is $\pm$~0.05~\AA~ and $\pm$~0.1 respectively. The indexing 
of the layers is relative to the topmost bulk Pd layer (0$^{th}$ Pd 
layer) for both Ni and Pd.
Each Pd layer spacing is always referred to the underneath Pd layer.
Each Ni layer spacing, but the 
1$^{st}$ one, is always referred to the underneath Ni layer.
 See text 
for the explanation of the layer filling behavior.}
\begin{ruledtabular}
\begin{tabular}{cccccc}
\multicolumn{3}{c}{{\bf 8~\AA~ film}}
&\multicolumn{3}{c}{{\bf 12~\AA~ film}}
\\
layer & spacing & filling & layer & spacing & filling \\ 
\hline
\\
           &      &      & 10$^{th}$ Ni & 1.53 & 0.3 \\
           &      &      & 9$^{th}$ Ni  & 1.52\footnotemark[1] & 0.7 \\
           &      &      & 8$^{th}$ Ni  & 1.52\footnotemark[1] & 0.9 \\
           &      &      & 7$^{th}$ Ni  & 1.52\footnotemark[1] & 0.9 \\
6$^{th}$ Ni& 1.56 & 0.3  & 6$^{th}$ Ni  & 1.52\footnotemark[1] & 1.0 \\
5$^{th}$ Ni& 1.52\footnotemark[1] & 0.7 & 5$^{th}$ Ni & 1.52\footnotemark[1] & 0.9 \\
4$^{th}$ Ni& 1.52\footnotemark[1] & 0.7 & 4$^{th}$ Ni & 1.52\footnotemark[1] & 0.8 \\
3$^{rd}$ Ni& 1.86 & 1.0  & 3$^{rd}$ Ni  & 1.82 & 0.7 \\
2$^{nd}$ Ni\footnotemark[2]& 1.83 & 0.8 & 2$^{nd}$ Ni\footnotemark[2] & 1.82 & 0.7 \\
2$^{nd}$ Pd\footnotemark[2]& 2.07 & 0.2 & 2$^{nd}$ Pd\footnotemark[2] & 2.05 & 0.2 \\
1$^{st}$ Ni\footnotemark[2]& 2.18 & 0.3 & 1$^{st}$ Ni\footnotemark[2] & 2.08 & 0.3 \\
1$^{st}$ Pd\footnotemark[2]& 1.94 & 0.7 & 1$^{st}$ Pd\footnotemark[2] & 1.90 & 0.7 \\
\end{tabular}
\end{ruledtabular}
\footnotetext[1]{fixed parameter, after Ref.~\onlinecite{rizzi}}
\footnotetext[2]{intermixed layers}
\end{table}

\section{Discussion}

The evolution of the Ni/Pd(100) film, as a function of the thickness, 
is described by the following sequence.
In the early deposition stage a couple of intermixed layers are 
formed with strict pseudomorphic structure. Further deposition leads 
to the formation of strained layers whose lattice cell is 
pseudomorphic to the lateral substrate lattice, but presents a perpendicular 
compression in agreement with the elastic force constant model.
This elastically strained structure builds up to a critical 
thickness of 12-16~\AA, corresponding to a maximum of 10 layers 
(including at least 2 intermixed layers at the substrate-film 
interface), not all of them equally occupied. Beyond this thickness, 
domains of bulk-like structure are formed and the whole film 
structure is gradually changed through a first order phase transition. Phase 
co-existence is clearly observed up to 18~\AA. The strain 
release at the phase transition is not complete and the bulk-like phase 
gradually relaxes to the orthomorphic fcc phase with increasing thickness.

The formation of a few deposit layers with a full pseudomorphic 
structure (as indicated by XPD scan in Fig.~\ref{fig5} and rod scan 
analysis in table~\ref{table1}) was also observed for very thin 
films (2-3 monolayers) of 
Fe on Cu$_{3}$Au(100) \cite{bruno,lin} and Ni on Cu(100).\cite{mueller}
The highly strained structure 
of these layers is probably stabilized by the intermixing and might 
be a general behavior of metal heteroepitaxy, whenever 
alloying or surface segregation is chemically favoured. 
In the present system and before the phase transition, the degree of 
intermixing remains constant, i.e. the concentration of substrate 
atoms is much smaller in the next deposited layers. 
From the 3rd layer 
on, a compressed vertical spacing is established, in agreement with 
the value predicted by the elastic theory for the 
given in-plane expansion, dictated by the substrate lattice.

The sequence of rod scans in 
Fig.~\ref{fig7} clearly shows that there exists a maximum 
number of Ni layers that can be stabilized into a pseudomorphic 
structure (10 layers, included the intermixed 
ones), thus defining an effective critical thickness for the 
structural phase transition. 
This configuration is established at a bulk equivalent 
thickness of 12~\AA. 
The fact that the maximum number of layers with pseudomorphic fct structure 
does not change throughout the transition, but simply the fct 
domains decrease their homogeneity and size, indicates that the 
transition takes place through the lateral growth of the Ni bulk-like 
domains 
at the expense of the pseudomorphic fct phase. 
Thus the Ni bulk-like domains are 
not floating on top of the pseudomorphic fct phase nor the 
transition simply proceeds from top to the bottom, rather, when  
formed, the Ni bulk-like domains involve locally all the film layers, 
down to the Ni/Pd interface. 
A possible model for the Ni bulk-like domains could be 
that of wedges, with the apex at the substrate-film interface, which expand 
at the expense of the pseudomorphic phase.
Since the phase transition involves three-dimensional domains, it 
must be of the first order type. No time evolution of the 
diffracted peaks has been observed at room temperature  on the time 
scale of the experiments (a few hours for the rod scans), which 
indicates that the transition is kinetically slowed down by the film 
defects.

At the critical thickness, the intermixing at the 
Ni/Pd interface (almost constant up to the critical thickness) 
increases strongly (as indicated by XPS analysis in Fig.~\ref{fig3}). 
It is very tempting to associate the increase of the intermixing with 
the propagation of a defect from the Ni/Pd interface (whose early 
formation might be witnessed by the incomplete layer filling of the 
buried Ni layers at the critical thickness), which drives 
the transformation of the pseudomorphic fct phase. 
The Ni/Pd lattice mismatch of $\sim$~10~\%, although not so small, 
might still allow the formation of an ordered pattern of misfit dislocations 
originated by the stacking faults at the 
interface, which would yield characteristic satellites of the Bragg's peaks 
in the radial scans.\cite{renaud}
While we have not checked by in-plane XRD for these feature on non-equivalent 
diffraction peaks other than the ($\overline{2}$~0~0) one, 
LEED pictures taken at the 
critical thickness were reported to display a Moir\'e pattern with $(10 \times 10)$ 
periodicity.\cite{rizzi} 
These patterns clearly show one order of extra spots decorating the 
integer peaks along the $\left\langle011\right\rangle$ substrate symmetry 
directions, i.e. the direction of close-compact atom rows on the fcc(100) 
surface. From comparison with the present data (in particular with 
the intermixing behaviour), we can say 
that the strain release indicated by the extra spots is not limited 
to the Ni film surface, rather the Moir\'e pattern is originated by 
defects extending down to the substrate. 
The Ni/Pd(100) system, where the Ni film undergoes a structural 
transition between two phases with the same face centered (100) surface symmetry, 
resembles the behavior of the Fe/Cu(100) system, 
where a more dramatic structural change, 
from an fcc(100) to a bcc(110) phase, takes place. In the latter case, 
the transition of the Fe film was described through the 
formation of shear planes (from the film surface down to the substrate) along  
close-compact atom rows, which separate domains of different 
structural phase, like for 
a martensitic phase transition.\cite{kalki}

Concerning the growth of the newly formed bulk-like Ni domains, a mechanism 
like that driving the decoherence in semiconductor heteroepitaxy can 
be envisaged. The domain walls, either shear planes or 
misfit dislocations, propagate laterally on the Pd substrate, 
leaving behind a strongly intermixed region. This picture is fully 
consistent with the increase of the
Ni/Pd intermixing, which is observed to continue beyond the critical thickness 
(as indicated both by XPS analysis and by the Ni/Pd interface width 
obtained by the XRR analysis, Fig.~\ref{fig1})

Finally the newly formed bulk-like Ni film is still strained at small thickness 
(see radial scan at 22~\AA~ in Fig.~\ref{fig6}) and the relaxation to 
its bulk fcc structure proceeds gradually with increasing thickness.
This gradual relaxation, witnessed by the radial scans of the in-plane 
lattice parameters, is also seen in the perpendicular layered 
structure of the film. In fact the simple bulk structural models used 
to simulate the XRR energy scans yields a fitting quality which 
clearly improves from the 22 to 30~\AA~ film (see Fig.~\ref{fig1}).
Such a behavior is not surprising, since it has been observed for 
other metal films. As an example, the bulk structure of Fe deposited 
on Cu$_{3}$Au(100) (where the 7\% lattice mismatch is even smaller 
than the present one) is not fully 
recovered even for film thickness of the order of hundred of 
monolayers.\cite{rochow,schirmer}

\section{Conclusions}

We studied the growth of Ni on Pd(100) by XPS (both core levels and VB), 
XPD, XRR and XRD (both in- and out-of-plane) 
techniques. In particular, we followed the structural evolution 
of the Ni films as a function of the thickness. 
After the formation of a couple of intermixed layers, (which preserve 
a perpendicular spacing close to the substrate one) a laterally 
pseudomorphic phase, with a perpendicular strain in agreement with
the elastic theory, takes place. This pseudomorphic fct structure is 
stable up to 10 layers. Further Ni deposition leads to a gradual structural 
transformation of the whole film into a Ni bulk-like fcc phase. This phase 
transition, accompanied by a strong increase of intermixing at the 
Ni/Pd interface, proceeds through the lateral growth of the Ni 
bulk-like
domains at the expense of the pseudomorphic fct phase. The latter domains 
preserve their layered structure (maximum number of layers and corresponding 
spacings), while shrinking. Residual strain is still observed in the 
Ni bulk-like phase at a thickness of 30~\AA.


\begin{thebibliography}{99}

\bibitem{tsao} for an overview of the thermodynamics of semiconductor 
heteroepitaxy see chapter {\bf 5} in {\it Materials Fundamentals of Molecular Beam 
Epitaxy} edited by J.Y. Tsao, Academic Press Inc., San Diego, CA, 1993.  

\bibitem{renaud} for an overview of semi-coherent metal/oxide growth 
see G. Renaud, Surf. Sci. Rep. {\bf 32}, 1 (1998).

\bibitem{opitz}  R. Opitz, S. L\"{o}bus, A. Thissen and R. Courths, Surf.
Sci. {\bf 370},  293 (1997).

\bibitem{feau}  O.S. Hernan, A.L. Vazquez de Parga, J.M. Gallego and R.
Miranda, Surf. Sci. {\bf 415},  106 (1998).

\bibitem{demiguel} J.J. de Miguel, J. Camarero, J. de la Figuera, 
J.E. Prieto and R. Miranda, 
in {\it Morphological organization in epitaxial growth and removal}, 
ed. by Z. Zhang and M.G. Lagally, (World Scientific Co., Singapore 
1998), p. 367.

\bibitem{moroni} E.G. Moroni, G. Kresse, J. Hafner, J. Furthm\"uller, 
Phys. Rev. B {\bf 56} (1997) 15629.

\bibitem{kalki} K. Kalki, D.D. Chambliss, K.E. Johnson, R.J. 
Wilson and S. Chiang, Phys. Rev. B {\bf 48}, 18344 (1993). 

\bibitem{fecu} S. M\"uller, P. Bayer, C. Reischl, K. Heinz, B. Feldmann, H. Zillgen 
and M. Wuttig, Phys. Rev. Lett. {\bf 74}, 765 (1995); J. Giergiel, J. 
Shen, J. Woltersdorf, A. Kirilyuk and J. Kirschner, Phys. Rev. B {\bf 
52}, 8528 (1995).

\bibitem{fadley} C.S. Fadley
{\it Basics Concepts of X-Ray Photoelectron Spectroscopy} in
{\it Electron Spectroscopy, Theory, Techniques and Applications}, 
C.R. Brundle 
and A.D. Baker Eds. (Pergamon Press,1978), Volume II, Chapter 1.

\bibitem{bruno}  F. Bruno, D. Cvetko, L. Floreano, R. Gotter, C. Mannori, L.
Mattera, R. Moroni, S. Prandi, S. Terreni, A. Verdini and M. Canepa, Appl.
Surf. Sci. {\bf 162-163}, (2000) 340; F. Bruno, S. Terreni, L. Floreano, 
A. Cossaro, D. Cvetko, P. Luches, L. Mattera, A. Morgante, R. Moroni, 
M. Repetto, A. Verdini, M. Canepa, Phys. Rev. B {\bf 66} (2002) 045402.

\bibitem{rizzi} G.A. Rizzi, M. Petukhov, M. Sambi, G. Granozzi, 
Surf. Sci. {\bf 522}, (2003) 1.

\bibitem{aloisa}  L. Floreano, {\it et al.}, Rev. of Sci. Inst. {\bf 70}, 
3855 (1999); R. Gotter,  {\it et al.}, Nucl. Instrum. Meth. Phys. Res. 
A {\bf 467-468} (2001) 1468;
An updated presentation of the beamline can be found at
\url{http: //www.tasc.infm.it/ tasc/ lds/ aloisa/ aloisa.html}.

\bibitem{vlieg} E. Vlieg, J. Appl. Cryst. {\bf 33} (2000) 401. The 
software is freely distributed at \url{http://www.esrf.fr/computing/scientific/}.

\bibitem{bruno2}  F. Bruno, L. Floreano, A. Verdini,
D. Cvetko, R. Gotter, A. Morgante, M. Canepa and S.Terreni, J. Elec. 
Spectrosc. Related Phenom. {\bf 127} (2002) 85, cond-mat/0204404.

\bibitem{todorova} M. Todorova, E. Lundgren, V. Blum, A. Mikkelsen, S. 
Gray, J. Gustafson, M. Borg, J. Rogal, K. Reuter, J.N. Andersen, M. 
Scheffler, Surf. Sci. {\bf 541} (2003) 101.

\bibitem{hillebrecht} F.U. Hillebrecht, J.C. Fuggle, P.A. Bennet and Z. 
Zolnierek, Phys. Rev. B {\bf 27} (1983) 2179.

\bibitem{lin}  M.-T. Lin, J. Shen, W. Kuch, H. Jenniches, M. Klaua, C.M.
Schneider and J. Kirschner, Surf. Sci. {\bf 410} 298 (1998).

\bibitem{mueller} S. M\"uller, B. Schulz, G. Kostka, M. Farle, K. 
Heinz and K. Baberschke, Surf. Sci. {\bf 364} (1996) 235.

\bibitem{rochow}  R. Rochow, C. Carbone, Th. Dodt, F.P. Johnen and E.
Kisker, Phys. Rev. B {\bf 41},  3426 (1990).

\bibitem{schirmer}  B. Schirmer, B. Feldmann and M. Wuttig, Phys. Rev. B. 
{\bf 58} 4984 (1998).

\end{thebibliography}
\end{document}